\documentclass[aps,pre,twocolumn,superscriptaddress,showpacs,letter,nobalancelastpage,citeautoscript]{revtex4}

\usepackage{epsfig,graphicx}

\begin{document}
\title{Intermolecular Effect in Molecular Electronics}

\author{Rui Liu}
\affiliation{Department of Chemistry, Duke University, Durham, North Carolina 27708-0354}

\author{San-Huang Ke}
\affiliation{Department of Chemistry, Duke University, Durham, North Carolina 27708-0354}
\affiliation{Department of Physics, Duke University, Durham, North Carolina 27708-0305}

\author{Harold U. Baranger}
\affiliation{Department of Physics, Duke University, Durham, North Carolina 27708-0305}

\author{Weitao Yang}
\affiliation{Department of Chemistry, Duke University, Durham, North Carolina 27708-0354}


\begin{abstract}
We investigate the effects of lateral interactions on the conductance of two molecules connected in parallel to semi-infinite leads. The method we use combines a Green function approach to quantum transport with density functional theory for the electronic properties. The system, modeled after a self-assembled monolayer, consists of benzylmercaptane molecules sandwiched between gold electrodes. We find that the conductance increases when intermolecular interaction comes into play. The source of this increase is the indirect interaction through the gold substrate rather than direct molecule-molecule interaction. A striking resonance is produced only 0.3 eV above the Fermi energy.
\end{abstract}



\maketitle
Modern molecular electronics began in 1974 when Aviram and Ratner\cite{aviram} proposed a rectifier based on an asymmetric molecular tunneling junction. The size of molecular electronic devices -- on the nanometer scale -- provides advantages in cost and efficiency, while synthetic modification of molecules may enable manipulation of their electrical properties. All these advantages promote the possible use of such systems in future molecular electronic technologies. In fact, several fundamental devices have recently been demonstrated \cite{rectifier1,resonant1,switch1,switch2}.  In many of these, self-assembled monolayer (SAM) systems were used as they are a controllable step toward single molecule electronics.

Molecules thiolated on a gold surface comprise a frequently studied SAM. For instance, scanning tunneling measurements show that conjugated molecules with one methylene group between the sulfur and aromatic ring have a commensurate $ ( \sqrt{3} \times \sqrt{3} ) R 30^{\circ} $ structure \cite{herring}. Sulfur is chosen because it has two valence electrons, allowing bonding to both molecule and surface. Among other possibilities, oxygen has high electronegativity leading to a high dipole barrier, while Se and Te result in lower conductance as suggested by calculations on a simple phenyl-dithiolate molecule \cite{SanHuang}. At high coverage, the SAM forms via the strong S-Au bond \cite{SAM1,SAM2,SAM3}; a strong chemical bond is more likely to generate stable, reproducible junctions and avoid a large charge-transfer induced barrier which may mask the electronic signature of the molecule.

The conductance of a SAM may involve both intra-molecular and inter-molecular transport.  In the case of densely packed conjugated molecules \cite{Cui1,Cui2,Kushmerick}, the lateral interaction through delocalized $\pi$-orbitals could mask the individual contribution of a presumably isolated molecule. Thus, when two molecules with conductance $G_{1}$ and $G_{2}$ are sandwiched in parallel between two leads, the resulting conductance $ G_{1,2} $ is not simply equal to $G_{1} + G_{2}$. Such considerations cast doubt on the common procedure whereby the conductance of a single molecule is deduced from $G_{\mathrm{total}}$ by simply dividing by the number of molecules attached to the leads. Conversely, theoretical calculations on a single molecule may not be relevant for comparison to SAM experiments. 

For parallel molecular wires that connect directly to the electrodes, indirect interaction was found by Yaliraki and Ratner \cite{interchain2}. They set the interchain hopping to zero, and saw an increase in conductance, in the absence of any coupling between two molecules. A linear superposition law has been found using a tight-binding description \cite{interchain3}. The intensity of coupling as a function of equilibrium site distance was taken as evidence for interaction through the Au surface. Lang and co-workers \cite{langinter} calculated the low bias conductance of cumulene on jellium as a function of molecule-molecule distance. The range of intermolecular distance considered, however, is 3.5-6.5 a.u. -- less than 9.4 a.u., the distance between the closest adsorption sites of Au (111) surface.

\begin{figure}[b]
\includegraphics[width=1.in,clip]{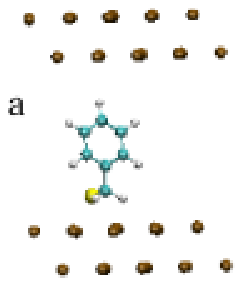}
\includegraphics[width=1.in,clip]{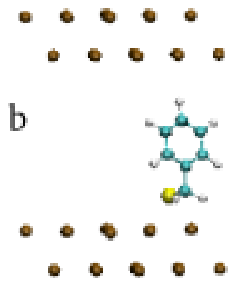}
\includegraphics[width=1.in,clip]{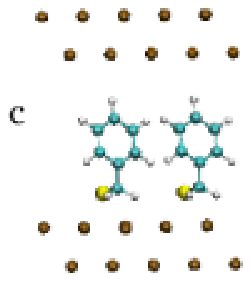}
\caption{\label{fig:brgstructure} (Color online) Structure of (a),(b) single benzylmercaptane molecules at each of two bridge sites on a Au(111) lead and (c) the corresponding double molecule system. Each Au layer contains 9 atoms. Brown, white, blue and yellow denote Au, H, C, and S atoms, respectively.}
\end{figure}

Since aromatic molecules are widely used in molecular electronics, we study the effect of intermolecular coupling on the conductance of two benzylmercaptanes (BM) sandwiched between semi-infinite gold leads. The geometry of the molecules (Fig.~\ref{fig:brgstructure}) is that of possible crystalline structures for the infinite SAM. The suggested herringbone packing of a BM-SAM \cite{herring} has two kinds of mutual structures: displaced $\pi$ stacking and vertical packing. Here we focus on the displaced $\pi$ stacking because of the shorter distance between delocalized $\pi$-bonds. Our density functional theory plus Green function approach has been described elsewhere \cite{San-Huang}. We used the Kohn-Sham method in the generalized gradient approximation for the electronic structure calculation, with numerical atomic orbitals (double zeta plus polarization) as the basis set, by means of the SIESTA program \cite{siesta}.

Fig.~\ref{fig:brgstructure} shows the structure of one or two BM moleclues adsorbed on the gold surface at bridge sites. The distance between the two equilibrium adsorption sites for the BM moelcules is 9.4 a.u. Unlike in a break junction \cite{MCB-junction}, each molecule has only one S-Au contact. A second contact for conductance measurements is provided by a STM tip contacting the top hydrogen atom. We start by considering single adsorbates in high symmetry sites on the Au(111) surface -- top, bridge, fcc hollow, and hcp hollow. According to our calculations, the hcp and top sites are not local minima; this agrees with methylthiolate results \cite{site3} showing that the hcp site is unstable at high coverage. As shown in Table \ref{tab:sitepreference}, the tilt angle of the relaxed fcc conformation is $7^{\circ}$ less than that of the relaxed bridge site. This implies more steric repulsion between the surface and the C-S back bond. In the end, the bridge site is favored by 0.16~eV.

\begin{table}[t]
\caption{\label{tab:sitepreference}The initial and relaxed site, the adsorption energy with respect to the fcc-bridge case, the height of the sulfur atom from the gold surface, the average of the three nearest neighbor Au-S distances, and the S-C bond length and angle $\theta$ to the surface normal. (81 $ k $-points are used, yielding a converged adsorption energy \cite{site3}.) }
\begin{tabular}{cccccccccc}
\hline\hline
Initial & Relaxed &$ \triangle E_{\mathrm{ad}}$(eV) & $h$ $(\AA)$ & $\overline{r}_{\mathrm{S-Au}} ( \AA) $&$ r_{\mathrm{s-c}} ( \AA ) $&$ \theta^{\circ} $\\
\hline
bridge&fcc-bridge&0&2.1&2.530&1.880&80\\
fcc&fcc&0.16&2.0&2.623&1.883&73\\
\hline\hline
\end{tabular}
\end{table}

\begin{table}[b]
\caption{\label{tab:intercoupling} Charge transfer from the leads to molecule and conductance of single and double molecules adsorbed at bridge or fcc sites. The subscript ``1'' denotes a single molecule at the central adsorption site, ``2'' a single molecule closer to the edge of the lead, and ``1,2'' two molecules. $\Delta Q$ is defined as $(Q_{1,2}-Q_{1}-Q_{2})/(Q_{1}+Q_{2})$ and $\Delta G$ similarly. The conductances are in units of $10^{-4} \times 2e^2/h$.}
\begin{tabular}{ccccccccc}
\hline\hline
Site&$Q_{1}$&$Q_{2}$&$Q_{1,2}$&$ \Delta Q$(\%)& $G_{1}$&$G_{2}$&$G_{1,2}$&$\Delta G$(\%)\\
\hline
$\mathrm{bridge}$ &0.202&0.071&0.266&$-5.5$ &2.6&1.7&\ 7.1&+65\\
$\mathrm{fcc}$ &0.199&0.110&0.292&$-2.5$ &3.5&7.1&16.8&+58\\
\hline\hline
\end{tabular}
\end{table}

The conductance and charge transfer changes due to intermolecular coupling are shown in Table \ref{tab:intercoupling}.  The small difference between $Q_{1}$ and $Q_{2}$, or $G_{1}$ and $G_{2}$, is due to the different position on the surface. We see that the change in charge transfer due to the insertion of a second molecule is very small. In contrast, the conductance increase is large, 65\% and 58\% for adsorption at the bridge and fcc sites, respectively. \textit{This demonstrates clearly that the conductance of a N-molecule SAM is not simply N times the conductance of a single molecule}.

The charge transfer is essentially a local issue because the sulfur atom does not connect directly to the conjugated aromatic ring. Therefore, the charge transferred to the sulfur cannot be efficiently shared with the core part of the molecule (aromatic ring). The methylene group between the sulfur and the aromatic ring does provide a support tool so that the BM will be straight up when adsorbed; on the other hand, it destroys the conjugate character of the molecule and so blocks the charge transfer. This trade-off between mechanical and electrical properties is a fundamental dilemma for molecular electronics: the inserted methylene group makes the molecule relatively insensitive to environmental changes -- crucial for reproducibility -- but its conductance is low.


\begin{figure}[b]
\includegraphics[width=3.in,clip]{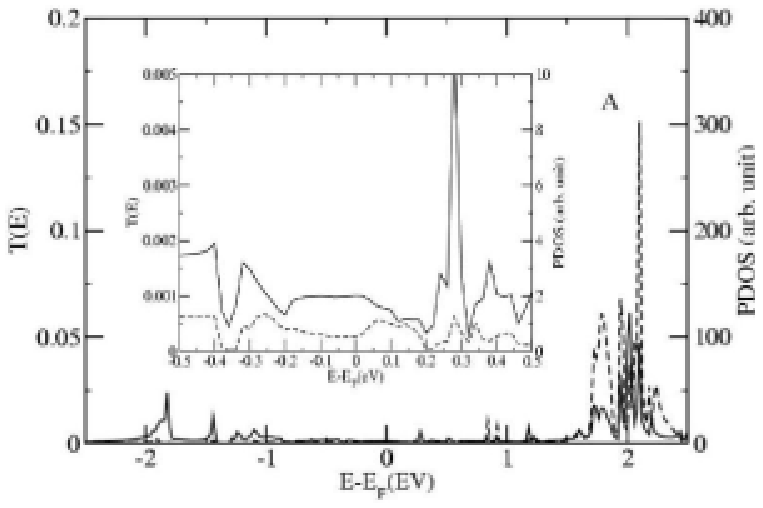}
\includegraphics[width=3.in,clip]{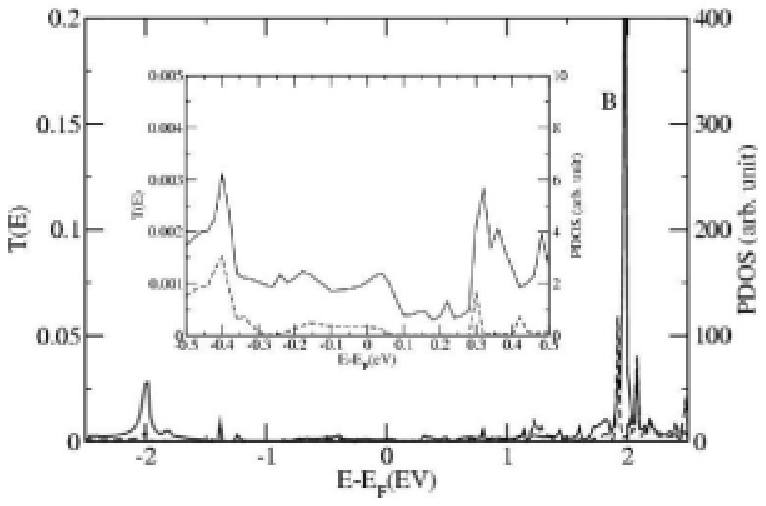}
\includegraphics[width=3.in,clip]{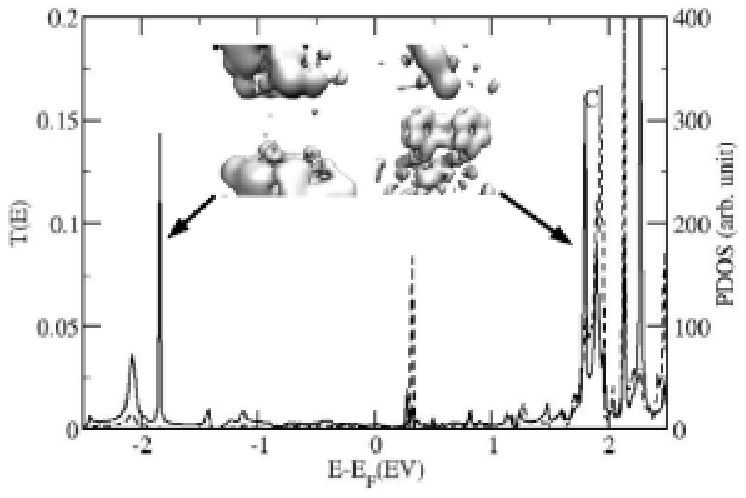}
\caption{\label{fig:brgplot} Transmission (dashed line) and PDOS (solid line) of BM molecuels adsorbed at bridge sites: (a) single molecule at site 1, (b) single at site 2, and (c) double molecule. The inset in (c) shows surfaces of constant local density of states (LDOS) at the HOMO (left) and LUMO (right) energies.
}
\end{figure}


\begin{figure}
\includegraphics[width=1.1in,clip]{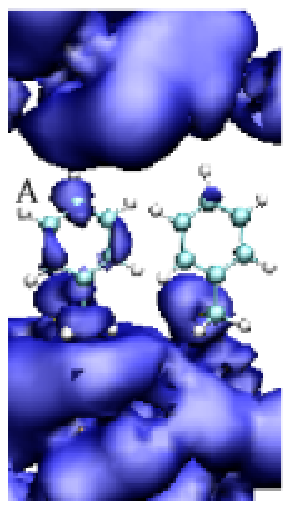}
\includegraphics[width=1.1in,clip]{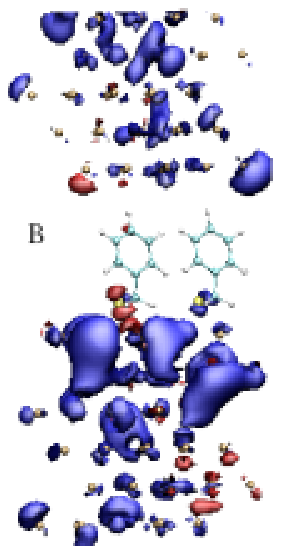}
\caption{\label{fig:peak} (Color online) (a) A surface of constant LDOS at 0.3 eV for the double molecule superposed on a ball-and-stick model. Note the involvement of $\pi$-orbitals on the central (site 1) BM. (b) Difference between LDOS of double molecule and that of single site-1 case. The blue and red denote positive and negative surface values, respectively. Brown, white, blue and yellow denote Au, H, C, and S atoms. (bridge site)}
\end{figure}

\begin{figure}
\includegraphics[width=2.5in,clip]{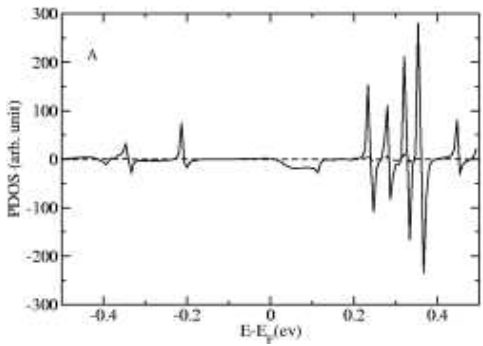}
\caption{\label{fig:brg-fcc-D-Pdos} 
Difference of DOS between double molecule case and sum of both single molecule cases projected on the aromatic rings (dashed) and the S-Au interface (solid).  For the latter, the S atom plus the two nearest layers of Au are used.  The change in PDOS near the Fermi energy clearly comes from the indirect interaction through the Au.
(bridge site)}
\end{figure}

\begin{figure}
\includegraphics[width=1.3in,clip]{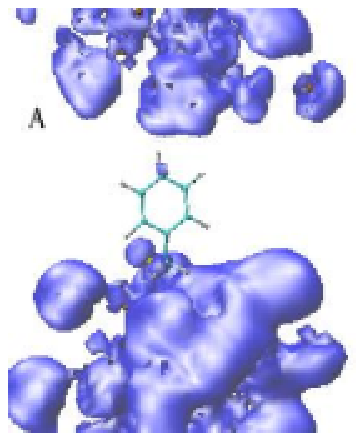}
\includegraphics[width=1.3in,clip]{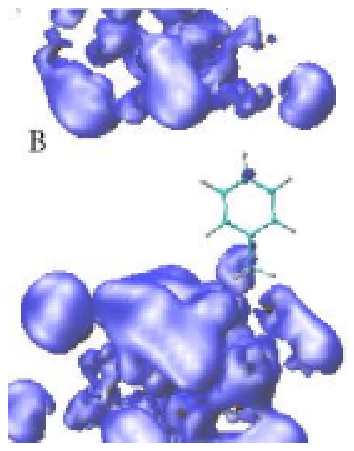}
\includegraphics[width=1.3in,clip]{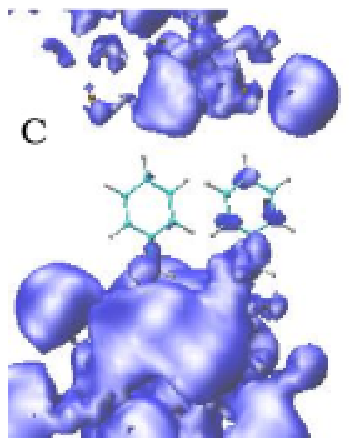}
\includegraphics[width=1.3in,clip]{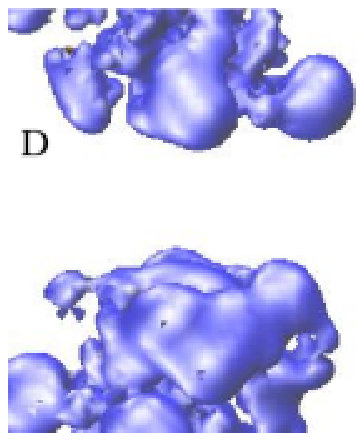}
\includegraphics[width=1.3in,clip]{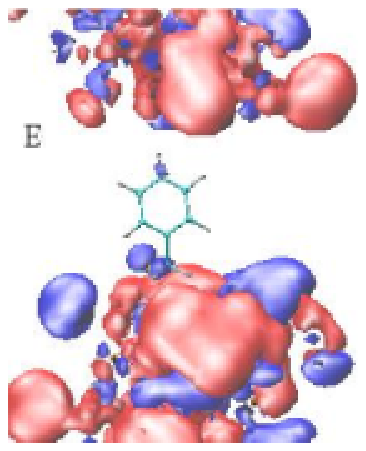}
\includegraphics[width=1.3in,clip]{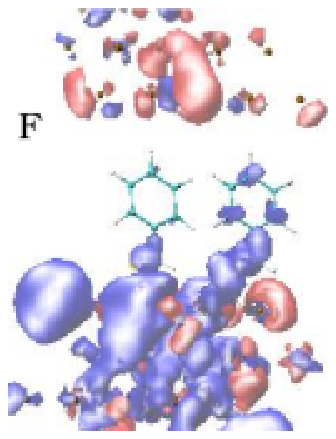}
\caption{\label{fig:LDOS} Surfaces of constant LDOS at the Fermi energy superposed on models of the molecules. (a) Single BM molecule at site 1. (b) Single molecule at site 2. (c) The Au lead without BM molecules. (d) Double molecule. (e) Difference between single molecule at site 1 and bare lead. (f) Difference between double molecule and bare lead. Note the electron deficient areas around the S-Au bonds; with two molecules present, this deficiency is decreased thus leading to better transport properties. Blue and red denote positive and negative values, respectively. Brown, white, blue and yellow denote Au, H, C, and S atoms, respectively. The energy window is $ \pm $0.02 eV around the Fermi level. (bridge site)}
\end{figure}

We now use the equilibrium transmission, $T(E)$, and the density of states projected on the BM molecule(s), PDOS, to further analyze our system; see Fig. \ref{fig:brgplot}. The coupling to the continuum of metal states leads, of course, to shifting and broadening of the molecular energy levels \cite{shiftbroaden1,shiftbroaden2}. Note that the curves for the two single molecule cases are very similar [panels (a) and (b)]; the difference comes only from different location with respect to the edges of the lead. The HOMO-LUMO gap is 3.8~eV in an isolated BM molecule calculation, and the same energy scale remains when connected to leads. From the inset of Fig.~\ref{fig:brgplot}(c), we see that the HOMO is localized at S atoms, which contribute low transmission. On the other hand, the LUMO is a delocalized $\pi$-orbital and leads to large $T$. In the gap, the transport process is non-resonant tunneling yielding low transmission.

Both $T(E)$ and the PDOS change upon going to a two molecule system (Fig. \ref{fig:brgplot}). A very striking feature appears: a narrow resonance about 0.3~eV above the Fermi energy. This feature appears in both the fcc and bridge site cases, but is not present for an isolated double BM system. Such a feature is relevant to molecular electronics since in the $I$-$V$ curve at low bias it would cause a region of negative differential conductance.

To show the nature of the resonance, we plot  in Fig.
Fig.~\ref{fig:peak}(a) a surface of constant local density of states (LDOS) for the energy 0.3 eV.  The resonance is clearly made from LUMO related anti-bonding $\pi$-orbitals on the central BM molecule.  The difference between the LDOS for the double molecule and that for the single site-1 case is shown in panel (b). An enhanced electron density appears at the sulfur connection when molecule 2 is added. The better S-Au coupling that this causes is apparently responsible for the resonance peak at 0.3 eV.  

Our results show clearly the large changes that interaction among the molecules of a SAM can induce.  Yet, the question remains as to
what brings about those alterations -- direct molecule-molecule interaction between, for instance, the aromatic rings, or indirect interaction through the gold surface? 


We answer this question by looking at how the DOS changes from the single molecule to double molecule cases. Fig.~\ref{fig:brg-fcc-D-Pdos} shows the change of the DOS projected on either the aromatic rings or the S-Au interface. Large fluctuations take place at the S-Au interface but not on the rings.  The flucutations occur both at energies near the edges of the HOMO-LUMO gap, which is natural from shifting of levels due to changed coupling, and near the Fermi energy.  \textit{Thus, the dominant interaction between the molecules is indirectly through the Au leads, and this controls the transport behavior} \cite{pdos-different-site}.  

To further understand the changes in conductance at the Fermi energy, we focus on the LDOS at that energy (see Fig.~\ref{fig:LDOS}).  
It is known, of course, that sulfur bonds well with nearby gold atoms, but we see only a small LDOS around the Fermi energy [panels (a)-(c)].  Thus, this type of bonding does \textit{not} provide a good conduction channel. This is an example of a structurally favorable anchoring group which provides reproducible measurements but is not a good choice for a transparent contact. The bottleneck of molecular electronic devices is, in fact, this connection between leads and molecule.  It depends largely on the choice of lead material and head group atoms. For better performance, the coupling should not only be strong far from Fermi energy. In our case the LDOS does not spread to the core molecule from the lead, because of either the large gap of the molecule or the methylene group blocking the conjugation of the system. 

Fig.~\ref{fig:LDOS}(e) tells us that when there is only one BM molecule, a density deficient region forms around the S-Au contact at the Fermi level, another demonstration of the poor conducting properties of this bond.  On the other hand, Fig.~\ref{fig:LDOS}(f) shows that when two molecules are present, there is an increase of LDOS in between them.  The S atom tends to push electrons at the Fermi energy away from it; however, when two molecules get together, the mutual pushing in some sense eliminates the electron deficient region caused by a single S atom, and so yields a better connection with increased conductance. 

Our calculations thus provide a concrete example of weakly interacting parallel molecules adsorbed on a gold surface, one of the scenarios investigated by Yaliraki and Ratner \cite{interchain2} with a model Hamiltonian. The direct molecule-molecule interaction can be neglected because of the large HOMO-LUMO gap and limiting states at the Fermi level. It is the indirect interaction through the gold substrate that modifies the conductance. Close packed molecules can change the electron density at the Fermi energy of gold leads near the surface, and this effect extends to more than four layers of gold atoms before decaying. Thus, a nearby molecule could have a dramatic effect on the conduction due to a modification of the LDOS of gold leads. 

This work was supported in part by the NSF (DMR-0103003).

\bibliography{inter}

\begin{thebibliography}{23}
\bibitem[{pdo()}]{pdos-different-site}
\bibinfo{note}{From the importance of the S-Au bond, one naturally infers that
  the change in PDOS as well as $T$ should be sensitive to the adsorption site.
  We see this for our fcc adsorption site calculations (not shown).}
\end{thebibliography}
\end{document}